\documentclass{ws-procs9x6}

\begin{document}

\title{Warped brane world supergravity,
flipping, and the Scherk-Schwarz mechanism}

\author{Zygmunt~Lalak and Rados\L aw~Matyszkiewicz}

\address{Institute of Theoretical Physics\\
University of Warsaw, Poland}


\maketitle

\abstracts{We demonstrate the relation between the Scherk-Schwarz mechanism and flipped 
gauged brane-bulk supergravities in five dimensions. 
We discuss the form of supersymmetry violating Scherk-Schwarz terms
in pure supergravity and in supergravity coupled to matter. 
Although the Lagrangian mass terms that arise as the result of the Scherk-Schwarz redefinition of fields are naturally of the order of the inverse radius of the orbifold, the effective 4d physical mass terms are rather set by the scale 
$\sqrt{|\bar{\Lambda}|}$, where $\bar{\Lambda}$ is the 4d cosmlogical constant.} 
\section{Introduction}%
The issue of hierarchical supersymmetry breakdown in supersymmetric 
brane worlds is one of the central issues in the quest for 
a phenomenologically viable extra-dimensional extension of the Standard Model.
Many attempts towards formulating scenarios of supersymmetry breakdown 
that use new features offered by extra-dimensional setup have been made 
\cite{Horava:1996vs,Antoniadis:1997ic,Dudas:1997jn,Nilles:1998sx,Lalak:1997zu,Lukas:1998tt,Ellis:1998dh,Falkowski:2001sq,Zucker:2000ks,Barbieri:2000vh,Bagger:2001ep}. One of them is supersymmetry breakdown triggered by imposing 
nontrivial boundary conditions on field configurations along the compact 
transverse dimensions usually referred to as the Scherk-Schwarz mechanism.
The initial investigation of this mechanism was based on the assumption 
that the 5d Minkowski vacuum is consistent with sources assumed to define the models under investigation, and, in consequence, (super)gravity wasn't playing 
any important role in these scenarios. However, it has been shown recently 
\cite{Falkowski:2001sq,Brax:2002vs,Lalak:2002kx} that the nontrivial supergravity 
background, consistent with sources, should be taken into account and 
may play an important role modifying the physics of the model.  
In fact, it is precisely  partial 
`unification' of the Standard Model with gravity that makes the Brane World 
scenarios so intriguing and appealing.
\vskip 0.5cm
{\small \noindent Talk given at the 1st International Conference on String Phenomenology, Oxford, UK, July 6 - 11, 2002.}
\newpage 
\noindent In particular,
the dynamical treatment of gravity is necessary when one discusses the 
stabilization of the extra dimension. 
 
Among other things we have found in \cite{Lalak:2002kx} that the simple flipped supergravity forms 
the locally supersymmetric extension of the $(++)$ bigravity model of 
Kogan et. al. \cite{Kogan:2000vb,Kogan:2000xc}. 
In such a setup one circumvents 
the van Dam-Veltman-Zakharov observation about the nondecoupling of 
the additional polarization states of the massive graviton. 
The size of the residual four-dimensional cosmological constant can be tuned 
to arbitralily small values by taking the distance between branes suitably 
large. Although the Lagrangian mass terms that arise as the result of the Scherk-Schwarz redefinition of fields are naturally of the order of the inverse radius of the orbifold, the effective 4d physical mass terms are rather set by the scale of the 4d cosmlogical constant.
\section{Flipped and detuned supergravity in five dimensions}
The simple N=2 d=5 supergravity multiplet contains metric tensor (represented by 
the vielbein $e^m_\alpha$), two gravitini $\Psi^A_\alpha$ and one vector field $A_\alpha$ -- the graviphoton. We shall consider gauging of a ($U(1)$) subgroup of the global $SU(2)_R$ symmetry of the 5d Lagrangian. In general, coupling of bulk fields to branes turns out to be related to the gauging, and the bulk-brane couplings will preserve only a subgroup of the $SU(2)_R$, see \cite{Brax:2001xf}. 
Covariant derivative contains both gravitational and gauge connections:
        \begin{equation} 
        D_\alpha\Psi_\beta^A=\nabla_\alpha\Psi_\beta^A+ A_\alpha P^A_B\Psi_\beta^B\ ,
        \end{equation} 
where $\nabla_\alpha$ denotes covariant derivative with respect to gravitational transformations and $ P= P_i \, {\rm i}\, \sigma^i$ is the 
gauge prepotential. The pair of gravitini satisfies symplectic Majorana condition
 $\bar{\Psi}^A\equiv\Psi_A^\dagger\gamma_0=(\epsilon^{AB}\Psi_B)^TC$ where $C$ is the charge conjugation matrix and $\epsilon^{AB}$ is antisymmetric $SU(2)_R$ metric (we use $\epsilon_{12}=\epsilon^{12}=1$ convention). 
If one puts $ P=0$ and stays on the circle, then as the twist matrix 
one may take any $SU(2)$ matrix acting on the symplectic indices $a=1,2$. 
On a circle the $U(1)$ prepotential takes the form $ P = g_S s_a \, i \, \sigma^a$ and the twist matrix is $U_\beta = e^{i\, \beta \, s_a \sigma^a}$. However, in this case the unbroken symmetry is a local one, and the Scherk-Schwarz condition is equivalent to putting in a nontrivial Wilson line \cite{Lalak:2002kx}.

When one moves over to an orbifold $S^1 / \Gamma$, one needs to define in addition to the gauging the action of the space group $\Gamma$ on the fields. 
Let us take $\Gamma = Z_2$ first. Then we have two fixed points at $y=0,\pi$, 
and we can define the action of $Z_2$ in terms of two independent boundary 
conditions ($\Psi$ stands here for a doublet of symplectic-Majorana spinors or for a doublet of scalars, like two complex scalars from the hypermultiplet)   
\begin{equation} \label{warunkiSS}
        \Psi(-y)={\hat{Q}}_0\Psi(y)\ ,\quad\Psi(\pi r_c-y)={\hat{Q}}_\pi\Psi(\pi r_c+y)\ ,
        \end{equation}
        where $\hat{Q}_0$, $\hat{Q}_\pi$ are some arbitrary matrices, independent of the space-time coordinates, such that  ${\hat{Q}}_0^2={\hat{Q}}_\pi^2=1$ . Conditions (\ref{warunkiSS}) imply:
        \begin{equation}
        \Psi(y+2\pi r_c)={\hat{Q}}_\pi {\hat{Q}}_0\Psi(y)\ .
        \end{equation}
Hence, if the boundary conditions at $y=0$ and $y=\pi r_c$ are different, 
one obtains twisted boundary conditions with $U_\beta ={\hat{Q}}_\pi {\hat{Q}}_0$.
It is easy to see that $U_\beta \hat{Q}_{0,\pi}  U_\beta= \hat{Q}_{0,\pi}$,
which is the consistency condition considered in \cite{Bagger:2001ep}.
This is immediately  generalized to $S^1/(Z_2 \times Z_2')$ with two fixed points 
for each of the $Z_2$s, $y=0,\frac{1}{2}\pi r_c, \pi r_c, \frac{3}{2} \pi r_c$, and independent 
$\hat{Q}_y$ at each of the fixed points.  

If one writes $\hat{Q}_\pi \hat{Q}_0=\exp(i\beta_a\sigma^a)$, the condition (\ref{warunkiSS}) is solved by
        \begin{equation}
        \Psi=e^{i\beta_a\sigma^a f(y)}\hat{\Psi}\ ,
        \end{equation}
        where $\hat{\Psi}$ is periodic on the circle and $f(y)$ obeys the conditions
        \begin{equation}
        f(y+2\pi r_c)=f(y)+1\ ,\quad f(-y)=- f(y)\ .
        \end{equation}
        
When one expresses the initial fields $\Psi$ through $\hat{\psi}$, the kinetic term in the Lagrangian generates  mass terms for {\em periodic} fields $\hat{\Psi}$:
        \begin{equation}
        \bar{\Psi}\gamma^M\partial_M\Psi\supset if'\bar{\hat{\Psi}}\gamma^5\beta_a\sigma^a\hat{\Psi} \ .
        \end{equation}
        \vskip0.3cm
%\end{document}
%\subsection{Scherk-Schwarz mechanism in the $SU(2)$ R-symmetry of 5d gauged 
%supergravity}
Let us now move on to the specific case of a 5d supergravity with a 
gauged $U(1)$ subgroup of the $SU(2)$ R symmetry. 
%We will consider case with fifth dimension as an orbifold, which is interesting f%rom Randall-Sundrum scenario point of view. 
The $Z_2$ action on the gravitino is defined as follows:
\begin{equation} 
\Psi^A_{\mu,5}(-y)=\pm\gamma_5(Q_0)^A_B\Psi^B_{\mu,5}(y)\ ,\:\:\Psi^A_{\mu,5}(\pi r_c-y)=\pm\gamma_5(Q_\pi)^A_B\Psi^B_{\mu,5}(\pi r_c+y)\ ,
\end{equation} 
%\begin{eqnarray} \label{gbcond}
 %       &\Psi^A_\mu(-y)=\gamma_5(Q_0)^A_B\Psi^B_\mu(y)\ ,\quad\Psi^A_5(-y)=-\gamma_5(Q_0)^A_B\Psi^B_5(y)\ ,&\nonumber\\&\Psi^A_\mu(\pi r_c-y)=\gamma_5(Q_\pi)^A_B\Psi^B_\mu(\pi r_c+y)\ ,\quad\Psi^A_5(\pi r_c-y)=-\gamma_5(Q_\pi)^A_B\Psi^B_5(\pi r_c+y),&
%\end{eqnarray}
and the parameters $\epsilon^A$ of the supersymmetry transformations obey the same boundary conditions as the 4d components of gravitini. 
%\end{document}
Symplectic Majorana condition 
($(Q_{0,\pi})^C=\sigma_2 (Q_{0,\pi})^{*}\sigma_2=-Q_{0,\pi}$) and normalization 
$(Q_{0,\pi})^2=1$ imply $Q_{0,\pi}=(q_{0,\pi})_a \sigma^a$, where $(q_{0,\pi})_a$ are real parameters.
%\end{document}
We would like to gauge a $U(1)$ subgroup of the global $SU(2)$. 
In the general case \cite{Brax:2001xf} we can choose the prepotential 
of the form
\begin{equation} 
\label{prep}
        P = g_R \epsilon(y) R + g_{S} S,
\end{equation}
%\end{document}
where $R=r_a i \sigma^a$ and  $S=s_a i\sigma^a$. On an orbifold 
$S^1/(Z_2 \times Z_2')$ the expression $\epsilon(y)R$ gets replaced by 
$\bar{R}(y)$ which is a pice-wise constant matrix with discontinuities (jumps) at the positions of the four branes. The basic relation between the boundary conditions and the prepotential comes from the requirement, that under supersymmetry variations the 
transformed gravitino $\psi^{A}_\alpha + \delta \psi^{A}_\alpha$ 
should obey the same boundary conditions as $\psi^{A}_\alpha$. Taking into account that the gauge field present in the supersymmetry transformation of the gravitini is that graviphoton, whose 4d part we choose to take $Z_2$-odd with respect to each brane (we need only $N=1$ supersymmetry on the branes), and the fifth component is always even, we obtain the relations  valid for any segment containing a pair of naighbouring fixed points
\begin{equation} 
\label{barel}
 [ Q_{0, \pi}, R  ] = 0, \;\;  \{  Q_{0, \pi}, S  \} = 0 \, .
\end{equation}
For nonzero $R$ this implies $Q_y$ proportional to $R$, i.e. 
$Q_{y}= \alpha \, ( i \, \sqrt{R^2})^{-1} \, R$ with $\alpha= \pm 1$. The simplest case  
of interest corresponds to $Q_0 = - Q_\pi $. As shown in \cite{Brax:2001xf}, in this case the closure of supersymmetry transformations reqires putting on the branes  equal tensions whose maginitude is determined by $R$
(we quote only the bosonic gravity part of the action):
%end{document}
\begin{equation}
S= -\int d^5 x \sqrt{-g_5} (\frac{1}{2}R + 6 k^2)+  6 \int d^5 x\sqrt{-g_4}k T (\delta(x^5) + \delta(x^5-\pi\rho))
\end{equation}
%end{document}
where $
k = \sqrt{ \frac{8}{9}(g_R^2 R^2+ g_S^2 S^2)}$ and 
%\end{equation}
% and
%\begin{equation}
$T= g_1\sqrt{\vec{R}^2}/\sqrt{(g_1^2 R^2+ g_2^2S^2)}$.
This is easily generalized. 
If on a $S^1/(\Pi\, Z_2)$ one takes boundary conditions given by pairs of $Q$ and $-Q$ one after another, then this implies that all branes 
on $S^1/Z_2$, $S^1/(Z_2 \times Z_2')$, $S^1/(\Pi\, Z_2)$ have the same brane tension. 
Assuming also $S \neq 0$ such a system gives a static vacuum with $AdS_4$ 
foliation and fixed radius of the orbifold. 
In the case of $Z_2$ the overall twist matrix is given by $U_\beta = - {1} $ and in the case of $Z_2 \times Z_2'$
there is no overall twist: $U_\beta= + {1}$. This may be
generalized again. From the analysis of \cite{Brax:2001xf} it follows that if 
in the boundary conditions $Q$ is followed by $+Q$ (and not $-Q$) on the next 
brane, then the brane tension on the second brane must be equal in magnitude but of opposite sign to that on the first brane. Together with the previous findings this leads to quasi-quiver diagrams where branes with brane tensions 
$\lambda$ 
and boundary conditions $(Q),(-Q),(Q),...,(-Q),(Q),(-Q)$
follow each other respecting 
$\Pi Z_2$ symmetry. 
These are 
locally supersymmetric backgrounds corresponding to the models of the type 
discussed in \cite{Barbieri:2000vh}.  

Let us discuss the genaration of the Scherk-Schwarz (nonsupersymmetric) 
mass terms in the case where { $Q_{0}$ and $Q_{\pi}$ are parallel.} Let us 
take for simplicity $Q_{0}=\sigma_3$. Then $Q_{\pi}=\alpha\sigma_3$, where $\alpha=\pm 1$, and the twisted boundary conditions take the form:
        \begin{equation}
        \Psi(y+2\pi r_c)=\alpha\Psi(y)\ .
        \end{equation} 
        For $\alpha=1$ we have usual case with periodic field. For $\alpha=-1$ we obtain `flipped' supersymmetry of \cite{Brax:2001xf}. Let us take a nonzero $S$-part of (\ref{prep}). 
Assume the prepotential of the form
        \begin{equation} 
        P=\frac{g}{\sqrt{2}}\left(\epsilon(y)\sigma_3+\sigma_1\right)\ .
        \end{equation} 
        For $\alpha=-1$ we can write:
        \begin{equation} 
        Q_{\pi} Q_0=-1=e^{i\beta(\epsilon(y)\sigma_3+\sigma_1)}\ ,
        \end{equation} 
        where $\beta=\pi+2k\pi$ and $k\in Z$. One obtains the following 
solution
        \begin{equation} 
        \Psi=e^{i\beta(\epsilon(y)\sigma_3+\sigma_1) f(y)}\hat{\Psi}\ ,
        \end{equation} 
        and supersymmetry violating mass terms, with the choice $f=y/(2 \pi r_c)$ are 
\begin{equation} 
-e_5 \frac{{1}}{2} \bar{\Psi}^A_\mu\gamma^{\mu \nu} \gamma^5
\frac{1}{2 r_c} ( \sigma^{3}_{AB} - \epsilon(y) \sigma^{1}_{AB})
\Psi_{\nu}^{B} - e_4 \delta(y-\pi r_c) \bar{\Psi}^{A}_\mu \gamma^{\mu\nu} 
\gamma^{\hat{5}}  \sigma^{1}_{AB} \Psi_{\nu}^{B}.
\end{equation} 
It is straightforward to conclude that the bulk Lagrangian after redefinition cannot be put into the form compatible with linearly realized supersymmetry. To see this, one should note that the only mass terms compatible with supersymmetry are given by  a prepotential. 
Supersymmetry requires, that the same prepotential determines the bulk scalar potential. In our case, we have redefined the gravitini only, hence the 
bulk potential term stays unchanged, independent of $\beta$. 
At the same time any prepotential that should describe the Scherk-Schwarz 
mass terms shall depend on $\beta$, hence the supersymmetric 
relation between 
mass terms and the scalar potential is necessarily violated. This is what we mean when we call the Scherk-Schwarz mass terms explicitly non-supersymmetric. 
On the other hand it is obvious, that the Scherk-Schwarz picture is 
equivalent indeed to a spontaneosly broken flipped supergravity.
The generalization to a quasi-quiver setup is obvious. 
One may notice, that only the bulk terms are proportional to the naive 
KK scale $1/r_c$. The scale of boundary terms is set by the 5d Planck scale. 

Let us note already here, that even though the symmetry that we are using 
to implement the Scherk-Schwarz mechanism may be a local one, the Scherk-Schwarz masses cannot be removed, as one may naively think, by means of a gauge transformation. Such a transformation would have to be a `large' one, leading from a periodic to an antiperiodic configuration. However, the definition of the model involves not only couplings in the Lagrangian but also the choice of specific boundary conditions. Hence such large gauge transformations connect two different (although physically equivalent) Hilbert spaces, and do not belong to the group of internal symmetries of our models.

%%%%%%%%%%%%%%%%%%%%%%%%%%%%%%%%%%%%%%%%%%%%%%%%%%%%%%%%%%%%%%%%%%%%%%%%%%%

\section{Wave functions and mass quantization in flipped supergravity}% (bi-supergravity)}

In this chapter we would like to have a closer look at the localization of wave functions and mass quantization in a simple model with twisted supersymmetry.
The specific twisted model we 
shall discuss here is the locally supersymmetric generalization of the 
$(++)$ bigravity model of \cite{Kogan:2000vb}.  
%        
%\subsection{$AdS_4$ compactification of the pure supergravity with 
%flipped boundary conditions (bi-supergravity)}
%
To procede let us go on to
the locally  supersymmetric model with a flip along the fifth dimension. 
The price for local supersymmetry and the trouble one encounters is 
the nonzero curvature in 4d sections. 
Let us take the supergravity action with the prepotential of the form: $P=g_{R}\epsilon(y){\rm i}\sigma_3R+g_{S}{\rm i}\sigma_1S$, and the brane action  required by supersymmetry: $
        S_{brane}=6\int d^5x\sqrt{-e_4}kT(\delta(y)+\delta(y-\pi r_c))$.
These sources do not admit the flat 4d Minkowski foliation, and 
the consistent solution is that of $AdS_4$ branes:
 $       ds^{2}=a^{2}(y)\bar{g}_{\mu\nu}dx^{\mu}dx^{\nu}+dy^{2}$, 
        where 
 $       a(y)=\sqrt{-\bar{\Lambda}}/k \;\cosh\left(k|y|-\frac{k\pi r_{c}}{2}\right)$, 
        and $\bar{g}_{\mu\nu}dx^{\mu}dx^{\nu}=\exp(-2\sqrt{-\bar{\Lambda}}x_{3})(-dt^{2}+dx^{2}_{1}+dx_{2}^{2})+dx_{3}^{2}$ is the four dimensional $AdS$ metric. 

The radius of the fifth dimension is determined by brane tensions:
$k\pi r_{c}=\ln (\frac{1+T}{1-T})$, and the  
normalization $a(0)=1$ leads to the 
fine tuning relation $\bar{\Lambda}=(T^{2}-1)k^{2}<0$. 
We are interested in small fluctuations around vacuum metric: $g_{\mu\nu}(x^{\rho},y)=a^{2}(y)\bar{g}_{\mu\nu}+\phi_{h}(y)h_{\mu\nu}(x^{\rho})$, where $h_{\mu\nu}(x^{\rho})$ is a 4d wave function in $AdS_{4}$  background ($(\Box_{AdS}+ 2\bar{\Lambda})h_{\mu\nu}=m^{2}h_{\mu\nu}$ \cite{Alonso-Alberca:2000ne}). 
        It is easy to check, that the massless mode $\phi_h=A_{0}\cosh^{2}(k|y|-k\pi r_{c}/2)$ satisfies the equation of motion in the bulk and the 
boundary conditions. 
Matching delta functions at fixed points leads to the mass quantization conditions which need to be solved numerically in a generic case. 
%        where we have introduced notation $t=\tanh(k\pi r_{c}/2)$ 
%and $c=\cosh^{-2}(k\pi r_{c}/2)$.
The equations of motion for gravitini are more troublesome, since 
the prepotential mixes $(\Psi_{\mu})_{1}$ and $(\Psi_{\mu})_{2}$ fields. 
As shown in \cite{Lalak:2002kx} one can eliminate this mixing and define modes that 
satisfy the four-dimensional Rarita-Schwinger equations in $AsS_4$, and 
are numbered by the $AdS_4$ mass $m$. 
The boundary conditions are imposed by the action of the        
$Z_2$ in the fermionic sector:
one needs to demand that the fields 
$(\Psi_{\mu})_2$ and $(\Psi_{\mu})_1$ vanish at the points $y=0$ and $y=\pi r_{c}$ respectively. 
This condition removes the mode $m=0$ from the spectrum. 
Again, finding the spectrum of the gravitini in a generic case requires 
numerical analysis. The general feature is that the mass spectrum  is visibly 
shifted with respect to the mass spectrum of the graviton. 

It turns out that one can compute analytically  the graviton and gravitini mass spectra in the limiting cases of a large extra dimension ($kr_{c} \gg 1$) and in the case of a small
extra dimension ($k r_c \ll 1$). In the regime $kr_{c} \gg 1$ we obtain the ultra-light graviton mode
        \begin{equation} 
        m^{2}_{light}\approx  12k^{2} e^{-k\pi r_{c}}\cosh^{-2}(k\pi r_{c}/2)\ ,
        \end{equation} 
and heavy modes
        \begin{equation} \label{lah}
          m^{2}_h \approx k^2 (-2+n+n^{2}) \cosh^{-2}(k\pi r_{c}/2)=(-2+n+n^{2})|\bar{\Lambda}|\ ,
        \end{equation}
for $n>1$. For gravitini we obtain: 
\begin{equation} \label{lag}
m^{2}_f \approx k^2  (n+1)^2 \cosh^{-2} (k\pi r_{c}/2)= (n+1)^2 |\bar{\Lambda}|\ .
\end{equation}
In the limit $k r_c \ll 1$ the equations give the following mass quantization
\begin{equation} \label{smah}
m^{2}_h=\frac{n^2}{r^{2}_c},\:\:\:  
m^{2}_{\psi}=\frac{1}{r^{2}_c}(\frac{1}{2}+n)^{2}.
\end{equation}
The approximate spectra for the gravitini masses that we have just obtained 
can be compared to the spectra of the massive spin-2 states belonging to the 
$AdS_4$
supermultiplets discussed earlier given the $AdS_4$ mass formula 
$m^2 = C_2 (E_0,s) - C_2(s+1,s)= E_0 (E_0 -3) -(s+1)(s-2)$ for representations 
$D(E_0, s)$. In the limit of dimensional reduction
this implies the 
spin-2 and spin-3/2 spectra $m^{2}_{2,n} = (E_0 + 1/2 +n)
(E_0 -5/2 +n)$ and $m^{2}_{3/2,n}= (E_0 +n)(E_0 + n -3) +5/4$, 
${m'}^{2}_{3/2,n}= (E_0 +n+1)(E_0 + n -2) +5/4$, 
for some $E_0$ and $n=0,1,2,...$ (in units of $\sqrt{-\bar{\Lambda}}$). 
The above mass formula fits the limiting ($k r_c \gg 1$) spectra of graviton 
(except the first massive mode) and gravitino masses 
(\ref{lah}) and (\ref{lag})
if $E_0 = 3/2$, but this value does not correspond to a unitary supermultiplet, since the necessary condition $E_0 > s+1$ \cite{Heidenreich:rz} 
is not fulfilled for $s=3/2$ and $E_0=3/2$. 
The natural value for dimensional reduction $5d \rightarrow 4d$ would be $E_0 =3$. This gives 
$m^{2}_{2,n}=n^2 + 4n + 7/4$, $m^{2}_{3/2,n}=(n+3/2)^2 -1$ and ${m'}^{2}_{3/2,n} = (n+5/2)^2 -5$,
again in clear mismatch with (\ref{lah}) and (\ref{lag}).
It is also clear that the graviton mass spectrum 
for a finite $k r_c$ differs from the supersymmetric one.  
In the case where the $r_c$ is much smaller than the curvature radius,
the spectrum of gravitons and gravitini approaches the usual, flat space, 
KK form with gravitini masses shifted with respect to these of the gravitons. 
Also in this limit the spectrum is clearly nonsupersymmetric, and the shift 
is due solely to the twisted boundary conditions.   

One can see that even in the limit $r_c \gg 1/k$ supersymmetry is not restored,
and the branes do not decouple like in the supersymmetric Randall-Sundrum 
case. The nondecoupling may also be seen from the shape of the wave 
functions of the massive modes, see \cite{Lalak:2002kx}. 

To summarize the discussion of the supersymmetry breakdown in the case of the flipped supergravity let us inspect the equation for the Killing spinors: 
          \begin{eqnarray} 
          &\left(\frac{a'}{a}+\frac{2\sqrt{2}}{3}g_{1}\epsilon(y)|R|\right)\epsilon_+^A+\frac{2\sqrt{2}}{3}\gamma_5g_{2}|S|(\sigma^1)^A_B\epsilon_-^B=0&\nonumber\\ &\left(\frac{a'}{a}-\frac{2\sqrt{2}}{3}g_{1}\epsilon(y)|R|\right)\epsilon_-^A+\frac{2\sqrt{2}}{3}\gamma_5g_{2}|S|(\sigma^1)^A_B\epsilon_+^B=0&\ ,
          \end{eqnarray}
where $\epsilon^{A}_{\pm} = 1/2 (\delta^{A}_{B} \pm \gamma_5 Q^{A}_{B}) \epsilon^{B}$. These 
equations result in the condition
          \begin{equation} 
          \left(\left(\frac{a^{\prime 2}}{a^{2}} - \frac{8}{9}g_{1}^{2}R^{2}\right)-\frac{8}{9}g_{2}^{2}S^{2}\right)\epsilon_{\pm}^A=0\ ,
          \end{equation} 
and together with Einstein equations this implies that for non-vanishing $S$ there are no nontrivial solutions of the Killing equation.
\section{Summary and conclusions}
We have shown that flipped and gauged five-dimensional supergravity 
is closely related to the Scherk-Schwarz mechanism of symmetry breakdown. 
In this case the Scherk-Schwarz redefinition of fields connects two phases of the model. 
One phase is such that supersymmetry is broken spontaneously, in the sense that there do not 
exist vacua preserving some of the supercharges. In fact, one cannot undo this breakdown in a continous way, since the choice of the projectors on both branes is a discrete one - one cannot 
deform continously $Q$ into $-Q$ within the model. In particular, in the limit $r_c \rightarrow 0$ 
all gravitini (and all supercharges) get projected away. In the second, Scherk-Schwarz phase,
linear supersymmetry is not realized explicitly in the Lagrangian, hence one finds susy breaking masses and potential terms in the bulk and/or on the branes.  
However, the physics of the two phases has to be the same, as they are 
related by a mere redefinition of variables.    

We have found that the simple flipped 5d supergravity is a supersymmetrization of the 
$(++)$ bigravity with two positive tension branes. In the limit of the large interbrane separation 
there exists a ultra-light massive graviton mode in addition to the exactly massless mode
(but there is no a nearly degenerate superpartner). 

As an example the Scherk-Schwarz terms for gauged supergravity coupled to bulk matter have been worked out. 
One can see that performing the Scherk-Schwarz redefinition 
of scalar fields, which my be identified for instance with the higgs-like 
fields in the observable sector, one can create a complicated scalar 
potential. However, it will always contain the same physics as the locally 
supersymmetric lagrangian in the spontaneously broken phase, which is usually 
much simpler to analyse. The same comment concerns the fermionic sector. 
The Scherk-Schwarz masses for matter fermions 
superficially look like terms breaking 
supersymmetry in a hard way (like quartic terms in the potential), but 
the equivalence to the spontaneously broken phase guarantees cancellation of 
dangerous divergencies.  
The fact that the Scherk-Schwarz masses for chiral fermions do not belong to a 
linearly realized 5d supersymmetry may be seen from the observation, that 
supersymmetric masses are defined by the geometry of the quaternionic manifold 
and by the Killing vectors $k^i$, none of which had changed under the 
Scherk-Schwarz redefinition. 
To summarize, the redefinitions have broken linear supersymmetry both in 
hipermultiplet and in gravity sectors.

In the class of models discussed here it is the $AdS_4$ background that appears naturally as a static solution of the equations of motion. However, firstly, there exist nearby time-dependent solutions 
leading to Robertson-Walker type cosmology on branes, and secondly, in more realistic models the gravitational background we have described shall be further perturbed by nontrivial gauge and matter 
sectors living on the branes.

%%%%%%%%%%%%%%%%%%%%%%%%%%%%%%%%%%%%%%%%%%%%%%%%%%%%%%%%%%%%%%%%
\section*{Acknowledgments}
\noindent
The authors thank P. Brax and S. Theisen for very helpful discussions. Z. L. thanks CERN Theory Division where part of this project was done.  
This work  was partially supported  by the EC Contract
HPRN-CT-2000-00152 for years 2000-2004, by the Polish State Committee for 
Scientific Research grant KBN 5 P03B 119 20 for years 2001-2002.


\begin{thebibliography}{99} 

%\cite{Horava:1996vs}
\bibitem{Horava:1996vs}
P.~Horava,
%``Gluino condensation in strongly coupled heterotic string theory,''
Phys.\ Rev.\ D {\bf 54} (1996) 7561
%[arXiv:hep-th/9608019].
%%CITATION = HEP-TH 9608019;%%

%\cite{Antoniadis:1997ic}
\bibitem{Antoniadis:1997ic}
I.~Antoniadis and M.~Quiros,
%``Supersymmetry breaking in M-theory and gaugino condensation,''
Nucl.\ Phys.\ B {\bf 505} (1997) 109
%[arXiv:hep-th/9705037].
%%CITATION = HEP-TH 9705037;%%

%\cite{Dudas:1997jn}
\bibitem{Dudas:1997jn}
E.~Dudas and C.~Grojean,
%``Four-dimensional M-theory and supersymmetry breaking,''
Nucl.\ Phys.\ B {\bf 507} (1997) 553
%[arXiv:hep-th/9704177].
%%CITATION = HEP-TH 9704177;%%

%\cite{Nilles:1998sx}
\bibitem{Nilles:1998sx}
H.~P.~Nilles, M.~Olechowski and M.~Yamaguchi,
%``Supersymmetry breakdown at a hidden wall,''
Nucl.\ Phys.\ B {\bf 530} (1998) 43
%[arXiv:hep-th/9801030].
%%CITATION = HEP-TH 9801030;%%

%\cite{Lalak:1997zu}
\bibitem{Lalak:1997zu}
Z.~Lalak and S.~Thomas,
%``Gaugino condensation, moduli potential and supersymmetry breaking in  M-theory models,''
Nucl.\ Phys.\ B {\bf 515} (1998) 55
%[arXiv:hep-th/9707223].
%%CITATION = HEP-TH 9707223;%%

%\cite{Lukas:1998tt}
\bibitem{Lukas:1998tt}
A.~Lukas, B.~A.~Ovrut, K.~S.~Stelle and D.~Waldram,
%``Heterotic M-theory in five dimensions,''
Nucl.\ Phys.\ B {\bf 552} (1999) 246
%[arXiv:hep-th/9806051].
%%CITATION = HEP-TH 9806051;%%

%\cite{Ellis:1998dh}
\bibitem{Ellis:1998dh}
J.~R.~Ellis, Z.~Lalak, S.~Pokorski and W.~Pokorski,
%``Five-dimensional aspects of M-theory dynamics and supersymmetry  breaking,''
Nucl.\ Phys.\ B {\bf 540} (1999) 149
%[arXiv:hep-ph/9805377].
%%CITATION = HEP-PH 9805377;%%

%\cite{Falkowski:2001sq}
\bibitem{Falkowski:2001sq}
A.~Falkowski, Z.~Lalak and S.~Pokorski,
%``Four dimensional supergravities from five dimensional brane worlds,''
Nucl.\ Phys.\ B {\bf 613} (2001) 189
%[arXiv:hep-th/0102145].
%%CITATION = HEP-TH 0102145;%%

%\cite{Zucker:2000ks}
\bibitem{Zucker:2000ks}
M.~Zucker,
%``Supersymmetric brane world scenarios from off-shell supergravity,''
Phys.\ Rev.\ D {\bf 64} (2001) 024024
%[arXiv:hep-th/0009083].
%%CITATION = HEP-TH 0009083;%%

%\cite{Barbieri:2000vh}
\bibitem{Barbieri:2000vh}
R.~Barbieri, L.~J.~Hall and Y.~Nomura,
%``A constrained standard model from a compact extra dimension,''
Phys.\ Rev.\ D {\bf 63} (2001) 105007
%[arXiv:hep-ph/0011311].
%%CITATION = HEP-PH 0011311;%%

%\cite{Bagger:2001ep}
\bibitem{Bagger:2001ep}
J.~Bagger, F.~Feruglio and F.~Zwirner,
%``Brane induced supersymmetry breaking,''
JHEP {\bf 0202} (2002) 010
%[arXiv:hep-th/0108010].
%%CITATION = HEP-TH 0108010;%%
      
%\cite{Brax:2002vs}
\bibitem{Brax:2002vs}
P.~Brax and Z.~Lalak,
%``Brane world supersymmetry, detuning, flipping and orbifolding,''
Acta Phys. Polon. B {\bf 33} (2002) 2399.
%%CITATION = HEP-TH 0207102;%%

%\cite{Lalak:2002kx}
\bibitem{Lalak:2002kx}Z.~Lalak and R.~Matyszkiewicz,
%``On Scherk-Schwarz mechanism in gauged five-dimensional supergravity and  on its relation to bigravity,''
arXiv:hep-th/0210053.
%%CITATION = HEP-TH 0210053;%%

%\cite{Kogan:2000vb}
\bibitem{Kogan:2000vb}
I.~I.~Kogan, S.~Mouslopoulos and A.~Papazoglou,
%``A new bigravity model with exclusively positive branes,''
Phys.\ Lett.\ B {\bf 501} (2001) 140
%[arXiv:hep-th/0011141].
%%CITATION = HEP-TH 0011141;%%

%\cite{Kogan:2000xc}
\bibitem{Kogan:2000xc}
I.~I.~Kogan, S.~Mouslopoulos, A.~Papazoglou and G.~G.~Ross,
%``Multi-brane worlds and modification of gravity at large scales,''
Nucl.\ Phys.\ B {\bf 595} (2001) 225
%[arXiv:hep-th/0006030].
%%CITATION = HEP-TH 0006030;%%

%\cite{Brax:2001xf}
\bibitem{Brax:2001xf}
P.~Brax, A.~Falkowski and Z.~Lalak,
%``Non-BPS branes of supersymmetric brane worlds,''
Phys.\ Lett.\ B {\bf 521} (2001) 105
%[arXiv:hep-th/0107257].
%%CITATION = HEP-TH 0107257;%%

%\cite{Alonso-Alberca:2000ne}
\bibitem{Alonso-Alberca:2000ne}
N.~Alonso-Alberca, P.~Meessen and T.~Ortin,
%``Supersymmetric brane-worlds,''
Phys.\ Lett.\ B {\bf 482} (2000) 400
%[arXiv:hep-th/0003248].
%%CITATION = HEP-TH 0003248;%%

%\cite{Heidenreich:rz}
\bibitem{Heidenreich:rz}
W.~Heidenreich,
%``All Linear Unitary Irreducible Representations Of De Sitter Supersymmetry With Positive Energy,''
Phys.\ Lett.\ B {\bf 110} (1982) 461.
%%CITATION = PHLTA,B110,461;%%

\end{thebibliography}
\end{document}